\documentclass[aps,prl,twocolumn,superscriptaddress,preprintnumbers]{revtex4-2}
\pdfoutput=1
\usepackage{amsmath}
\usepackage{graphicx}
\usepackage{bbm}
\usepackage{amssymb}
\usepackage{enumitem}
\usepackage{booktabs}
\usepackage{verbatim} 
\usepackage{soul}
\usepackage{mathrsfs}
\usepackage[linktocpage=true,
colorlinks=true,
urlcolor=blue,
anchorcolor=blue,
citecolor=blue,
filecolor=blue,
linkcolor=blue,
menucolor=blue,
]{hyperref}

\newcommand{\ex}{\mathrm{e}}
\newcommand{\diff}{\mathrm{d}}

\newcommand{\R}{\mathbb{R}}

\newcommand{\C}{\mathbb{C}}

\newcommand{\hook}{\mathbin{\rule[.2ex]{.4em}{.03em}\rule[.2ex]{.03em}{.9ex}}}

\usepackage{color}

\def\nn{\nonumber}
\newcommand{\ii}{\mathrm{i}}

\newcommand{\Z}{\mathbb{Z}}

\newcommand{\fg}{g}

\newcommand{\coeff}{\xi}
\newcommand{\Fgrav}{F_{\mathrm{grav}}}
\newcommand{\sigmanew}{\sigma}
\newcommand{\sigmamass}{\sigma}
\newcommand{\Fvac}{F_{S^3}}

\begin{document}

\title{
Toric gravitational instantons in gauged supergravity}

\author{Pietro Benetti Genolini}
\affiliation{D\'epartment de Physique Th\'eorique, Universit\'e de Gen\`eve, 24 quai Ernest-Ansermet, 1211 Gen\`eve, Suisse}

\author{Jerome P. Gauntlett}
\affiliation{Blackett Laboratory, Imperial College, Prince Consort Road, London, SW7 2AZ, U.K.}

\author{Yusheng Jiao}
\affiliation{Blackett Laboratory, Imperial College, Prince Consort Road, London, SW7 2AZ, U.K.}

\author{Alice L\"uscher}
\affiliation{Mathematical Institute, University of Oxford, Woodstock Road, Oxford, OX2 6GG, U.K.}

\author{James Sparks}
\affiliation{Mathematical Institute, University of Oxford, Woodstock Road, Oxford, OX2 6GG, U.K.}

\begin{abstract}
\noindent We introduce a general class of 
toric gravitational instantons in $D=4$, $\mathcal{N}=2$ gauged 
supergravity, namely Euclidean supersymmetric solutions with $U(1)^2$ isometry. 
Such solutions are specified by a ``supergravity labelled polytope", 
where the labels encode the 4-manifold topology, the choice 
of magnetic fluxes, and certain signs associated with the Killing spinor. 
Equivariant localization allows us to 
write down the gravitational free energy for such a solution, 
assuming it exists, and study its properties. These results open the way for a systematic study 
of holography in this setting, where the dual large $N$ field theories 
are defined on the boundary 3-manifolds, which 
are (squashed) lens spaces $L(p,q)$ or generalizations
with orbifold singularities.
\end{abstract}

\maketitle

\section{Introduction}\label{sec:intro}

Gravitational instantons were introduced as Euclidean
solutions to the vacuum $D=4$ Einstein equations, possibly with a cosmological constant \cite{Hawking:1976jb,Gibbons:1978tef}.
They have been extensively studied both as saddle points in the context of the Euclidean path integral approach to quantum gravity
and also for their intrinsic geometric interest. In the context of the AdS/CFT correspondence it is natural to enlarge this 
perspective and consider gravitational instantons to also include Euclidean solutions to the equations of motion of supergravity theories
that asymptotically approach, in a suitable sense, hyperbolic space $H^D$. Via AdS/CFT, such solutions in $D$ dimensions
have a dual interpretation as CFTs on the Euclidean $(D-1)$-dimensional conformal boundary, with a curved metric and possible additional deformations, including magnetic fluxes and mass deformations, depending on the precise asymptotic fall off of the supergravity fields
as one approaches the boundary. Of particular interest are such solutions that preserve supersymmetry, i.e. admit solutions to the appropriate
Killing spinor equations.

Here we are interested in supersymmetric gravitational instantons of $D=4$, $\mathcal{N}=2$ Euclidean gauged supergravity coupled to 
$n$ vector multiplets. This theory is obtained by a Wick rotation of the Lorentzian theory, whose bosonic field content consists of a metric,
$n+1$ gauge fields and $n$ complex scalar fields.
We assume the theory has an $H^4$ vacuum solution with vanishing gauge fields and scalars, which is 
dual to a 
3-dimensional
$\mathcal{N}=2$ SCFT~\footnote{This is certainly the case when
the gauged supergravity theory arises from a consistent KK truncation, but we believe our results have more general 
applicability as commented upon
in the discussion section.}.
Of significant interest is the evaluation of the gravitational free energy, $\Fgrav$, which is simply the on-shell action of the gravitational instanton
evaluated with appropriate boundary terms, as this corresponds to the partition function of the dual SCFT via $Z=\text{exp}[-\Fgrav]$.
Supersymmetry implies that the solutions admit an R-symmetry Killing vector, which can be constructed as a bilinear from the Killing spinor.
Building on \cite{BenettiGenolini:2019jdz}, it has been recently shown \cite{BenettiGenolini:2023kxp,BenettiGenolini:2024xeo,BenettiGenolini:2024lbj}
that one does not require the explicit solution to evaluate $\Fgrav$.
Instead, a knowledge of the topology of the manifold and the R-symmetry vector is sufficient and one can use the
BVAB localization theorem \cite{BV:1982, Atiyah:1984px}
to express $\Fgrav$ in terms of the fixed point set of the R-symmetry in the interior of the solution, which consists of
isolated fixed points and/or fixed surfaces, called ``nuts" and ``bolts", respectively \cite{Gibbons:1979xm}.

Here we further develop this formalism for a sub-class of solutions which possess an additional Abelian isometry, and hence are ``toric",  
allowing us to be more
explicit. In particular, we use the results of \cite{OrlikRaymond} that capture the toric data of the geometry in terms of a labelled polytope;
this data has also appeared in the analysis of ordinary gravitational instantons with toric  symmetry 
\cite{Chen:2010zu} where it is referred to as a ``rod structure'' (for related work see \cite{Emparan:2001wk,Harmark:2004rm,Harmark:2005vn,Hollands:2007qf,Hollands:2008fm}).
Here we show that a suitably generalized ``supergravity labelled polytope" enables one to compute $\Fgrav$, provided that
the supersymmetric solution exists. Some explicit solutions are known, either for minimal gauged supergravity or for the STU model,
and we can easily recover known results for $\Fgrav$. However, the power of our formalism is that it applies to solutions that are unlikely to be ever constructed in closed form. Focusing on the STU model, this leads to concrete predictions for new saddle points associated with ABJM theory
on squashed spheres and lens spaces, which, in principle, could be verified using localization on the field theory side.  We first focus on 
smooth gravitational instantons, but the formalism is easily adapted to incorporate orbifold singularities, both in the bulk and on the boundary, including considering the dual SCFT on the product of a spindle with a circle, as well as on more general branched lens spaces.

\section{Toric 4-manifolds}\label{sec:toric}

We consider a class of non-compact 4-manifolds $M$
with a $T^2=U(1)^2$ action, using the description 
in \cite{OrlikRaymond,Calderbank:2002gy}. 
We assume $M$ is simply-connected and then the quotient space $P=M/T^2$ is a polygon. Here the $T^2$ acts freely on the pre-image of the open 
interior of $P$, while the boundary $\partial P$ is the image 
of points in $M$ that are fixed under various subgroups of $T^2$, 
as described below. 

We label the edges of $\partial P$ by an index $a=0,\ldots,d+1$, and order them so that edges $a=0$, and $a=d+1$ are non-compact, while the remainder 
are compact. 
 Attached to each edge is a
coprime pair of integers $\vec{v}_a\in \Z^2$, specifying the circle subgroup $U(1)\subset T^2$
that fixes a corresponding $T^2$-invariant 2-manifold $D_a\subset M$. 
Here we may introduce vector fields $\partial_{\psi_i}$, $i=1,2,$ that generate the $T^2=U(1)^2$ action, 
where $\psi_i$ are local coordinates with period $2\pi$, and write \footnote{Notice that $\vec{v}_a$ and $-\vec{v}_a$ specify the same $U(1)$ subgroup.}
\begin{align}\label{varphia}
\partial_{\varphi_a} = \sum_{i=1}^2 v_a^i \partial_{\psi_i}\, .
\end{align}
The vector field $\partial_{\varphi_a}$ generates the $U(1)\subset T^2$ 
fixing $D_a$.
When $M$ is a smooth manifold we necessarily have $D_0\cong D_{d+1}\cong \R^2$ 
while $D_a\cong S^2$
 for $a=1,\ldots,d$. 

Adjacent edges in $\partial P$ intersect at points $x_a=D_a\cap D_{a+1}$, $a=0,\ldots,d$, 
which correspond to isolated fixed points of the $T^2$ action. 
One can verify that $x_a$ is a smooth point of $M$ 
only if $\det (\vec{v}_a,\vec{v}_{a+1})=\pm 1$, and 
we use the sign ambiguity in specifying each $\vec{v}_a$ 
to fix the choice $\det (\vec{v}_a,\vec{v}_{a+1})=1$, for each 
$a=0,\ldots,d$.  Using the freedom to make 
$SL(2,\Z)$ transformations of the basis for $T^2$ 
we may also choose $\vec{v}_0=(-1,0)$.  
Finally, it will be convenient to introduce the 
intersection numbers, $D_{ab}\in\Z$, of 2-spheres $D_a$ and $D_b$, with 
\begin{align}\label{Dab}
D_{ab} = \begin{cases}\, 1 & b=a\pm 1\\
\, -\det(\vec{v}_{a-1},\vec{v}_{a+1})& b= a\\ 
\, 0& \mbox{otherwise}\end{cases}\, ,
\end{align}
for $1\leq a,b \leq d$. We also define $D_{10}=D_{d\, d+1}=1$.

\begin{figure}[ht!]
\begin{center}
\includegraphics[scale=.27]{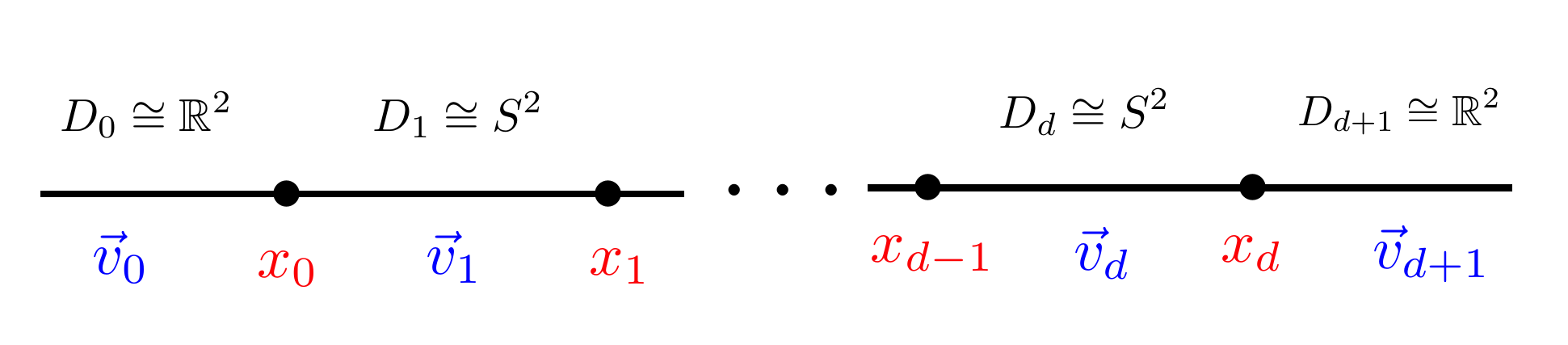}~
\caption{Toric data for a general $M$. The polytope $P$ 
may be taken to be the upper half-plane, with boundary data 
on $\partial P$ along the $x$-axis as shown.
}
\label{troicdatafig}
\end{center}
\end{figure}

\begin{figure}[ht!]
\begin{center}
\includegraphics[scale=.21]{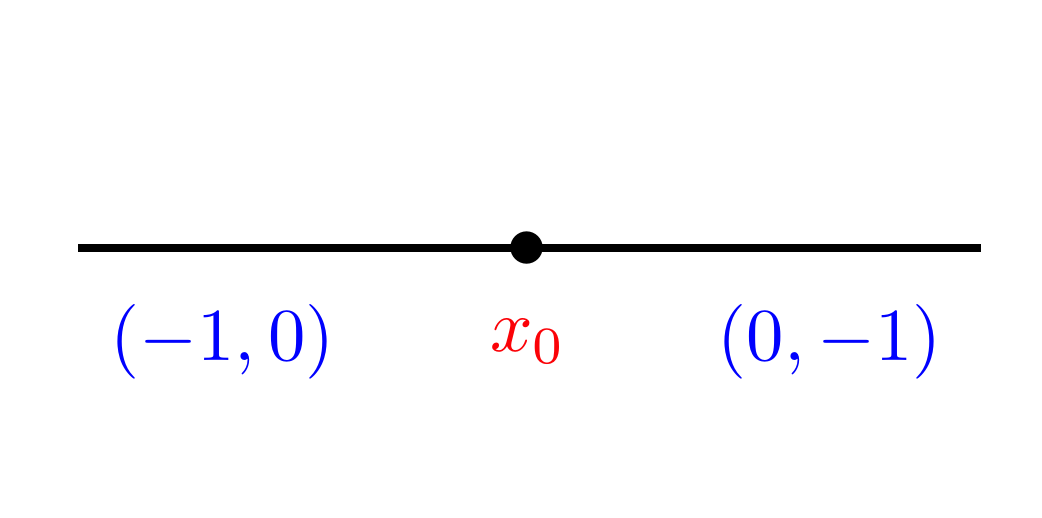}\quad
\includegraphics[scale=.21]{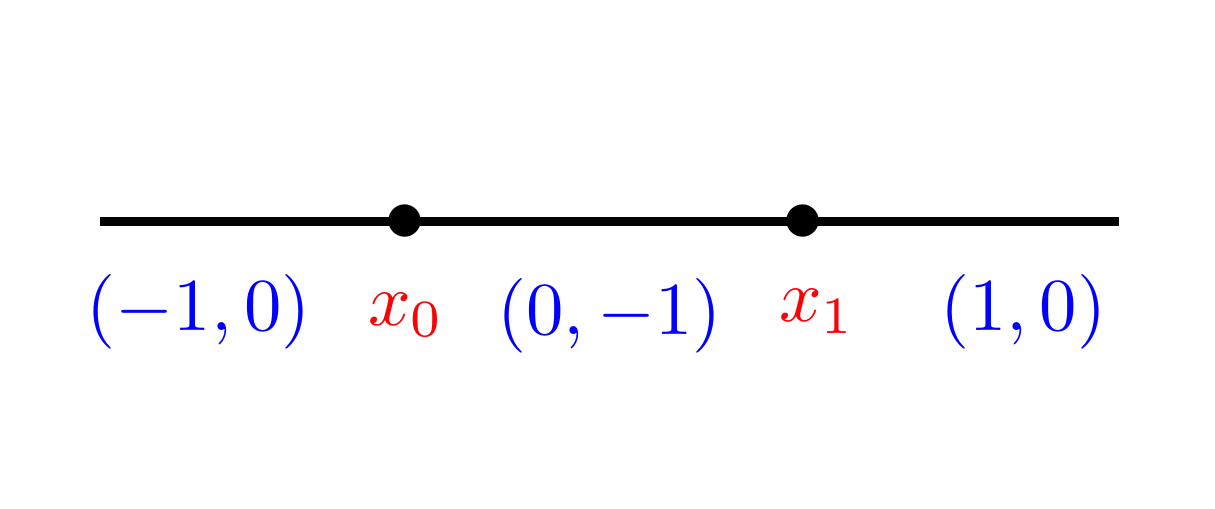}\quad
\caption{Toric data for $\mathbb{R}^4$ (left) and $\mathbb{R}^2\times S^2$ (right).
}
\label{troicdatafigexamples}
\end{center}
\end{figure}

This ``toric data'' may be summarized as in Fig.~\ref{troicdatafig}. 
We note that when $M$ admits a complex/symplectic structure that is compatible 
with the $T^2$ action, making $M$ a 
toric complex/symplectic manifold, 
then $P$ is also naturally convex, and the 
$\vec{v}_a$ may be interpreted as outward-pointing 
normal vectors to each edge. More generally there is no such convexity property. Labelled polytopes for $M=\R^4$ and 
$M=\R^2\times S^2$ are shown in Fig.~\ref{troicdatafigexamples}. 

A general toric vector field $\xi$ (i.e. one with flows generating a subgroup of 
$T^2$)
 may be written
\begin{align}\label{xibasis}
\xi = \sum_{i=1}^2 \coeff_i\mskip1mu  \partial_{\psi_i}\, ,
\end{align}
with coefficients $\coeff_i\in\R$. The tangent space to 
the fixed point $x_a\in M$ is $T_{x_a}\cong \R^2_{a,1}\oplus 
\R^2_{a,2}$, and the weights of $\xi$ on each 
$\R^2_{a,i}$ factor are $b_{a}^i\in \R$ with
\begin{align}\label{bweights}
b_{a}^1 = -\det (\vec{v}_{a+1},\vec{\xi}\mskip2mu)\, , \quad b_{a}^2 = \det (\vec{v}_a,\vec{\xi}\mskip2mu) \, .
\end{align}
Here $\vec{\xi}=
(\xi_1,\xi_2)$, with components as in \eqref{xibasis}. We can also write the weights as 
$b_a^i={\vec{u}}_a^{i}\cdot \vec \xi$, where ${\vec{u}}_a^{i}\in\Z^2$ for 
each $i=1,2$; 
specifically, $(u_a^{1})^i=-\epsilon^{ij}v^j_{a+1}$ and $(u_a^{2})^i=\epsilon^{ij}v^j_{a}$.
Notice that $b_a^2=-b_{a-1}^1$ holds for $a=1,\ldots,d$, 
or equivalently $\vec{u}_a^{2}=-\vec{u}_{a-1}^{1}$. 
The $\vec{u}_a^{i}$ may be interpreted as edge vectors, pointing inward towards $x_a$. 

\section{Supergravity solutions on \texorpdfstring{$M$}{M}}

We consider a subclass of supersymmetric solutions of Euclidean $D=4$, $\mathcal{N}=2$ gauged 
supergravity, coupled to $n$ vector multiplets with Fayet--Iliopoulos (FI) gauging.
The class of solutions are defined on a 4-manifold $M$ and are equipped with a real metric, real Abelian gauge fields $A^I$, $2n$ real scalar fields, and a Dirac spinor $\epsilon$ satisfying a Killing spinor equation. These comprise
a consistent subclass of the full space of solutions of the Euclidean supergravity, where \textit{a priori} the bosonic fields are complex and there are two independent Killing spinors \cite{Freedman:2013oja}. A more detailed discussion of the class of solutions that we study is presented in \cite{BenettiGenolini:2024lbj}.

We take the 4-manifold $M$ to be toric. The class of solutions we are studying have additional global data \cite{BenettiGenolini:2024lbj}, which we now define.
Firstly,
supersymmetric solutions have
an R-symmetry Killing vector \cite{BenettiGenolini:2024xeo,BenettiGenolini:2024lbj}
\begin{align}
\xi \equiv -\ii\epsilon^\dagger \gamma^\mu \gamma_5\epsilon\mskip1mu \partial_\mu\, .
\end{align}
Here $\epsilon$ is the Dirac Killing spinor, $\gamma^\mu$ 
are Hermitian generators of $\mathrm{Cliff}(4,0)$, 
and $\gamma_5\equiv \gamma^{1234}$ is the chirality operator.
We assume $\xi$ lies in the $T^2$ isometry group
of $M$, as in \eqref{xibasis}. 

Secondly,
the theory has real Abelian
gauge fields $A^I$, with $I=0,\ldots,n$ labelling the
graviphoton in the gravity multiplet together with the $n$ gauge fields in the vector multiplets. 
Each of these 
has an associated
 equivariantly 
closed form 
\begin{align}
\Phi^I \equiv F^I + \Lambda^I\, ,\qquad \diff_\xi \Phi^I=0\,,
\end{align}
where $\diff_\xi \equiv \diff - \xi\hook\, $. 
Here  $F^I=\diff A^I$ are the gauge field strengths and
$\Lambda^I$ are functions that depend on 
the scalar fields in the vector multiplets and certain scalar bilinears 
in the Killing spinor \cite{BenettiGenolini:2024xeo}, but their 
precise form will not be needed here. 
We also define the R-symmetry gauge field $A_R$ 
using the 
FI gauging parameters, $\zeta_I\in \mathbb{R}$, via
\begin{align}\label{defargf}
A_R\equiv \frac{1}{2}\zeta_I A^I\,.
\end{align}
For each vector field we must specify 
its magnetic flux through a basis of 2-cycles. 
These are provided by the 2-spheres $D_a$, 
 $a=1,\ldots,d$, where note that the two poles of $D_a$ are 
  $x_a$, $x_{a-1}$, 
with $b_a^2$ and $b_{a-1}^1=-b_a^2$ being the corresponding 
weights of $\xi$ on the tangent spaces to each pole in $D_a$, respectively.
We may then use the 
 BVAB localization formula 
 to compute the 
corresponding magnetic 
fluxes $\mathfrak{p}_a^I$ for $a=1,\ldots,d$:
\begin{align}\label{pflux}
\mathfrak{p}_a^I & \equiv \frac{1}{4\pi}\int_{D_a}F^I = \frac{1}{4\pi}\cdot 2\pi \left(
\frac{\Lambda^I_a}{b^2_a} + \frac{\Lambda^I_{a-1}}{b^1_{a-1}}\right) \nonumber\\ 
& = 
\frac{1}{2b^2_a}\left(\Lambda^I_a-\Lambda^I_{a-1}\right)\, ,
\end{align}
where $\Lambda^I_a\equiv \Lambda^I|_{x_a}$, $a=0,\dots, d$, are constants.

Thirdly, the Killing spinor of the supersymmetric solution, $\epsilon$, is
charged with respect to $A_R$ defined in \eqref{defargf}. It is also 
charged with respect to the Lie derivative $\mathcal{L}_\xi \epsilon = 
\xi^\mu\nabla_\mu \epsilon + \frac{1}{8}(\diff\xi^\flat)_{\mu\nu}\gamma^{\mu\nu}\epsilon$,  
and this leads to additional global data. In general the 
charge is dependent on the gauge chosen for $A_R$, but at a fixed point where $\xi=0$ and 
in a \emph{non-singular} gauge at that point we have~\cite{BenettiGenolini:2024xeo,BenettiGenolini:2024lbj}
\begin{align}\label{epsiloncharge}
\mathcal{L}_\xi \epsilon\mskip2mu\big|_{\mathrm{fixed}} = \frac{1}{8}\diff\xi^\flat_{\mu\nu}\gamma^{\mu\nu}\epsilon\mskip2mu\big|_{\mathrm{fixed}} = 
\frac{\ii}{4}\Lambda\mskip1mu \epsilon \mskip2mu\big|_{\mathrm{fixed}}\, .
\end{align}
Here $\Lambda\equiv \zeta_I \Lambda^I$. 

The left hand side 
of \eqref{epsiloncharge} may be computed as follows. 
Consider a particular 2-manifold $D_a$, where recall that 
$\partial_{\varphi_a}$ given by \eqref{varphia} 
rotates the normal bundle $ND_a$ to $D_a$ in $M$, 
acting on a copy of a normal $\R^2$ fibre with weight 1. 
The  Killing spinor $\epsilon\mskip2mu |_{D_a}$ 
will decompose as a tensor product of a 
spinor on $D_a$ with a spinor on the normal space 
$\R^2$. Since we assume that $T^2$ is an isometry, 
a short argument shows that $\epsilon\mskip2mu |_{D_a}$ 
must have definite charge under $\partial_{\varphi_a}$, given by 
 \begin{align}\label{sigmaa}
\mathcal{L}_{\partial_{\varphi_a}} \epsilon\mskip2mu \big|_{D_a} = \frac{\ii}{2}\sigmanew_a \epsilon \mskip2mu \big|_{D_a}\, .
\end{align}
Moreover, 
 a non-singular spinor $\eta$ 
on $\R^2$ of definite charge necessarily has 
$\mathcal{L}_{\partial_{\varphi_a}} \eta =\frac{\ii}{2}\sigmanew_a \eta =\pm  \frac{\ii}{2} \eta$, correlated with its chirality, which fixes each  
 $\sigmanew_a=\pm1$. 

We now look at a fixed point $x_a=D_a\cap D_{a+1}$. 
Observe that we may also write the R-symmetry vector \eqref{xibasis} as
\begin{align}\label{xichangebasis}
\xi = b_a^1 \mskip2mu\partial_{\varphi_a} + b_a^2\mskip2mu\partial_{\varphi_{a+1}}
\, ,
\end{align}
so that $\partial_{\varphi_a}$, $\partial_{\varphi_{a+1}}$ 
form the natural basis for $T^2$ acting on $T_{x_a}\cong \R^2_{a,1}\oplus 
\R^2_{a,2}$. From \eqref{epsiloncharge}, \eqref{sigmaa} we then 
obtain
\begin{align}\label{Lambdaa}
\Lambda_a \equiv \Lambda\mskip2mu |_{x_a} = 2(\sigmanew_a b_a^1 +\sigmanew_{a+1}b_a^2)\, .
\end{align}

From the second equality in \eqref{epsiloncharge} we may instead 
characterize the $\sigmanew_a$ by the projection conditions
\begin{align}
\ii\gamma^{12}\epsilon\mskip2mu |_{x_a} = \sigmanew_a \epsilon\mskip2mu |_{x_a}\, , \quad 
\ii\gamma^{34}\epsilon\mskip2mu |_{x_a} = \sigmanew_{a+1} \epsilon\mskip2mu |_{x_a}\, ,
\end{align}
where the local orthonormal frame at $x_a$ 
is such that $(\mathrm{e}^{1},\mathrm{e}^2)$ and 
$(\mathrm{e}^{3},\mathrm{e}^4)$ form a basis for 
$\R^2_{a,1}$ and 
$\R^2_{a,2}$, respectively. 
It follows that 
\begin{align}
\gamma_5\epsilon\mskip2mu |_{x_a} = \chi_a \epsilon\mskip2mu |_{x_a}\, ,
\end{align}
and the chirality of $\epsilon$ at $x_a$ is 
\begin{align}
\chi_a = -\sigmanew_a\sigmanew_{a+1}\in \{\pm 1\}\,.
\end{align}

Finally, as in \cite{BenettiGenolini:2019jdz} we note that the magnetic flux for the R-symmetry
\eqref{defargf} is fixed via 
\begin{align}\label{Rflux}
\zeta_I\mathfrak{p}_a^I & = \frac{1}{2b_a^2}\left(\Lambda_a - \Lambda_{a-1}\right)  = 
\sum_{b=0}^{d+1} D_{ab}\sigmanew_b \nonumber \\ & 
= \sigmanew_{a-1}+\sigmanew_{a+1}+D_{aa}\sigmanew_a\, .
\end{align}
Here the second equality makes use of equation \eqref{Lambdaa} together with the regularity conditions $\det (\vec{v}_{a-1},\vec{v}_{a}) =1=\det (\vec{v}_a,\vec{v}_{a+1})$,  
and we have used \eqref{Dab}. 

\section{Boundary data}

Fixing $\vec{v}_0=(-1,0)$, in general we may write 
$\vec{v}_{d+1}=(q,-p)\in \Z^2$ 
with $p$ and $q$ coprime 
integers, which we take to be non-negative. The residual $SL(2,\Z)$ transformations that stabilize 
$\vec{v}_0$  shift $q\mapsto q + b\mskip2mu p$, for $b\in\Z$, and using 
this freedom we may fix $0<q<p$, unless $p=1$, in which case $q=0$, 
or $p=0$, in which case $q=1$. 
The boundary $\partial M$ of the 4-manifold $M$
is then the lens space 
\begin{align}
\partial M = L(p,q)=S^3/\Z_{p}\, ,
\end{align} 
where the $\Z_{p}$ acts on $\C^2\supset S^3$ via 
$(z_1,z_2)\mapsto (\omega_{p}\mskip1mu z_1,\omega_{p}^{q}\mskip1mu 
z_2)$, with $\omega_{p}=\ex^{2\pi \ii/p}$ a primitive $p$'th root of unity. 
Here when $p=0$ this is understood to be $L(0,1)=S^1\times S^2$. 
The choice of toric 4-manifold $M$, with 
labelled polytope data $\vec{v}_a$, $a=1,\ldots,d$, 
may then be viewed as a \emph{resolution} of the 
orbifold singularity $\C^2/\Z_{p}$ (see \cite{Calderbank:2002gy}), or as a smooth filling of the lens space boundary. 
The latter is the more holographic perspective.

Having fixed a choice of $\partial M$, in holography
 we must also fix other UV data on $\partial M$. Similarly to 
\eqref{pflux}, for the non-compact $D_0\cong \R^2\cong  D_{d+1}$ we may 
define 
\begin{align}\label{Delta}
\Delta^I_{S} \equiv \frac{1}{4\pi}\int_{D_{0}}F^I \, , \quad 
\Delta^I_{N} \equiv \frac{1}{4\pi}\int_{D_{d+1}}F^I \, .
\end{align}
We then  introduce 
\begin{align}\label{ys}
y^I_{S} & \equiv \frac{1}{2b^2_0}\Lambda^I_0\, , \qquad 
\sigmamass^I_S \equiv -\frac{\ii}{4\pi} \Lambda^I|_{D_{0}\cap \mskip2mu\partial M}\,,
\nonumber\\
y^I_{N} & \equiv  \frac{1}{2b^1_d}\Lambda^I_d\,, \qquad 
\sigmamass^I_N \equiv -\frac{\ii}{4\pi} \Lambda^I|_{D_{d+1}\cap\mskip2mu \partial M}\,,
\end{align}
so that applying Stokes' theorem to  \eqref{Delta} leads to 
\begin{align}\label{UVIR}
\Delta^I_{S,N} + \ii\beta_{S,N}\sigmamass^I_{S,N} = y^I_{S,N}\, .
\end{align}
Here we have defined $\beta_S\equiv 2\pi/b^2_0$, $\beta_N\equiv 2\pi/b^1_d$. Geometrically these are the periods 
$\Delta\psi_{S,N}=\beta_{S,N}$, where 
we write $\xi |_{D_0} = \partial_{\psi_S}$, $\xi |_{D_{d+1}} = \partial_{\psi_N}$. Notice from \eqref{Lambdaa} that we have the constraints 
\begin{align}\label{yRconstraint}
\zeta_Iy^I_S = \sigmanew_1 + \sigmanew_0\mskip1mu\varepsilon_S\,,
\quad
 \ \zeta_Iy^I_N = \sigmanew_d + \sigmanew_{d+1}\varepsilon_N\,, 
\end{align}
where we have defined
\begin{align}\label{epsnsdef}
\varepsilon_S\equiv\frac{b_0^1}{b_0^2}\,,\qquad \varepsilon_N\equiv\frac{b_d^2}{b_d^1}\,.
\end{align}

On each copy of $D_0\cong \R^2\cong D_{d+1}$ we 
may write $F^I=\diff A^I_{S,N}$ globally, respectively,
 and then 
also choose a gauge for each integral in \eqref{Delta} where $A^I_{S,N}$ is non-singular 
at the origins $x_0\in D_0$, $x_d\in D_{d+1}$, respectively. This leads to the
 formulae $\Delta^I_S = \frac{1}{4\pi} \int_{D_0 \cap\mskip2mu \partial M} A^I_S$, 
$\Delta^I_N = \frac{1}{4\pi} \int_{D_{d+1} \cap\mskip2mu \partial M} A^I_N$. 
The parameters $\sigma^I_{S,N}$ are associated with mass deformations of the boundary SCFT
and also depend on the boundary geometry \cite{BenettiGenolini:2024lbj}.
Equation \eqref{UVIR} is then a ``UV-IR'' relation, 
where the left hand side depends on data on the UV 
boundary $\partial M$, while the right hand side 
only depends on fixed point data in the interior of $M$.  
Note, however, there is not necessarily a \emph{global} 
gauge choice on $M$ that allows this purely UV 
interpretation of both $\Delta^I_{S,N}$. 
Finally, note that the UV data $\Delta^I_{S,N}$, $\sigma^I_{S,N}$ are all constants.
While there certainly could be additional non-constant UV data \footnote{For example if the boundary 
was $S^2\times S^1$, the gauge fields could have holonomy around the $S^1$ that depend on the
position on $S^2$. Similarly we can have mass deformations that depend on all of the boundary coordinates.}, it is only this constant data which
enters the free energy. 

\section{Free energy}

We may now summarize the holographic problem, and specify the free parameters.
We fix a lens space boundary $\partial M=L(p,q)$, 
with an arbitrary $U(1)^2$ invariant metric,
and a toric 4-manifold $M$ that fills it with R-symmetry Killing vector $\vec\xi$ \eqref{xibasis}. $M$ is specified by a 
labelled polytope $P$, as in Fig. \ref{troicdatafig}, with 
$\vec{v}_0=(-1,0)$,
$\vec{v}_{d+1}=(q,-p)$. 
We have $d+1=\chi(M)$ fixed points/vertices $x_a$, $a=0,\ldots,d$, and 
a given solution will have
corresponding constant values $\Lambda^I_a$ of $\Lambda^I$ at those 
fixed points. On the other hand, fixing a choice of $y^I_S$
satisfying the constraint \eqref{yRconstraint}, 
one can solve the equations \eqref{pflux} for $\Lambda^I_a$ in terms 
of $y^I_S$, the $\mathfrak{p}^I_a$, and the toric data, for 
each $a=1,\ldots,d$: 
\begin{align}\label{lamprelation}
\frac{\Lambda_a^I}{2}&=b_0^2 y_S^I +\sum_{a'=1}^a b_{a'}^2\mathfrak{p}^I_{a'}\,.
\end{align}
In particular, we have
\begin{align}\label{yIN}
y^I_N &= \frac{b_0^2}{b_d^1} y^I_S + \sum_{a=1}^d \frac{b_a^2}{b_d^1}\mathfrak{p}^I_a\, ,
\end{align}
and this automatically satisfies the constraint in \eqref{yRconstraint}.

\begin{figure}[ht!]
\begin{center}
\includegraphics[scale=.27]{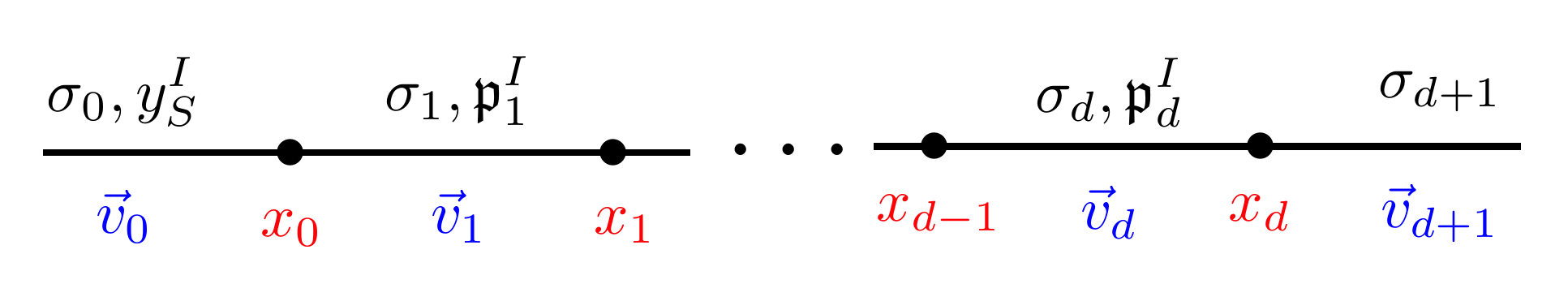}~
\caption{Supergravity labelled polytope. 
}
\label{troicdatafigdressed}
\end{center}
\end{figure}

A supersymmetric supergravity solution, assuming it exists, will therefore be specified by a \emph{supergravity labelled polytope}, 
as in Fig.~\ref{troicdatafigdressed}, where in 
addition to the toric data in Fig. \ref{troicdatafig} we specify 
magnetic fluxes $\mathfrak{p}^I_a$ for each internal $D_a\cong S^2$,
$a=1,\ldots,d$, and also specify chirality data $\sigmanew_a=\pm 1$ 
for each $a=0,\ldots,d+1$. The magnetic fluxes 
must satisfy the R-symmetry constraint \eqref{Rflux}, 
but this allows
 a freedom to shift a given choice  $\mathfrak{p}^I_a\mapsto \mathfrak{p}^I_a +
\mathfrak{f}^I_a$
by \emph{flavour magnetic 
fluxes} $\mathfrak{f}^I_a$, where by definition 
$\zeta_I \mathfrak{f}^I_a=0$. 
The gravitational free energy for such a solution is \footnote{To obtain this result for the holographically renormalized free energy, supersymmetry requires specific finite counterterms as well as a Legendre transform to 
implement alternative quantization on half of the scalar fields \cite{BenettiGenolini:2024lbj}. Interestingly, in carrying out this procedure one finds
that the contribution to the free energy from the conformal boundary then vanishes \cite{BenettiGenolini:2024lbj}.}, for a generic choice of 
$\vec{\xi}$ where all fixed points are the isolated vertices $x_a$, given by~\cite{BenettiGenolini:2024xeo,BenettiGenolini:2024lbj}
\begin{align}\label{Fgravsimp}
&F_{\mathrm{grav}} 
= -\frac{\pi}{G_N}\sum_{a=0}^d \frac{\chi_a}{4b_a^1b_a^2}\ii \mathcal{F}(\Lambda^I_a) \,.
\end{align}
Recall here that $\chi_a=-\sigmanew_a\sigmanew_{a+1}\in\{\pm 1\}$ is the chirality of the Killing
spinor at the fixed point $x_a$, $G_N$ is the Newton 
constant, and $\mathcal{F}$ denotes the prepotential of the theory. 
Using \eqref{ys}, \eqref{lamprelation} we also have
\begin{align}\label{Fgravcute}
	F_{\mathrm{grav}} 
	= -\frac{\pi}{G_N}&\Bigg[\frac{\chi_S}{\varepsilon_S}\ii \mathcal{F}(y^I_S)+\frac{\chi_N}{\varepsilon_N}\ii \mathcal{F}(y^I_N) \nn \\ &+\sum_{a=1}^{d-1} \frac{\chi_a}{b_a^1 b_a^2}\ii \mathcal{F}\Big(b_0^2 y_S^I +\sum_{a'=1}^a b_{a'}^2\mathfrak{p}^I_{a'}\Big)\Bigg]\, ,
\end{align}
where the $y_N^I$ are defined through \eqref{yIN}. 
For polytopes with $d=0$ or $d=1$ the above expression only includes the first term or the first
line, respectively.
It is now apparent that 
this free energy only depends on the supergravity labelled polytope, 
and is a function of the choice 
of R-symmetry vector $\vec{\xi}$ as in \eqref{xichangebasis} and choice of $y^I_S$
satisfying \eqref{yRconstraint}. 
With the latter constraint the $y^I_S$ are then $n$ degrees of freedom.
The UV-IR relation \eqref{UVIR} allows us to interpret 
both $\vec{\xi}$ and $y^I_S$ as UV data on the conformal boundary 
$\partial M$. In field theory one might have thought that 
$y^I_N=\Delta^I_N+\ii\beta_N \sigma_N$ could be specified entirely independently, but interestingly this is constrained 
to satisfy \eqref{yIN} by the bulk. 
The extension of these results to solutions 
where the fixed point set also includes 2-sphere bolts is discussed in
the supplementary material.

\section{Examples}\label{sec:examples}

In the following we illustrate our general formula \eqref{Fgravcute} on three examples with $d=0$, $1$ and $2$ respectively. While we keep the prepotential general, one can obtain results for the STU model and dual ABJM theory, by picking
\begin{align}\label{FSTU}
	\mathcal{F}(X^I) = -2\ii \sqrt{X^0 X^1 X^2 X^3}\, ,
\end{align}
and $\zeta_I=1$ for all $I=0,1,2,3$. 
We can also obtain results for minimal gauged supergravity, {\it e.g.} by setting the vector multiplets to zero,
then $I$ just takes the value 0 with
$\zeta_0 = 4$, and $\mathcal{F}(X^0) = -2 \ii (X^0)^2$.
In our conventions 
$\frac{\pi}{2G_N}=\Fvac$ is the free energy of the dual theory
on the round $S^3$ in the large $N$ limit.

\subsection{Example 1: \texorpdfstring{$M = \R^4$}{M=R4}}

Using the toric data
for $\R^4$ in Fig.~\ref{troicdatafigexamples}, with $d=0$, we straightforwardly compute 
\begin{align}\label{freeR4}
F_{\mathrm{grav}} = -2\frac{\chi}{\varepsilon_S}\ii\mathcal{F}(y_S^I) \Fvac\, ,
\end{align}
where $\varepsilon_S 
={\xi_1}/{\xi_2}$ and $y_S^I$ is constrained via \eqref{yRconstraint}, $\zeta_Iy_S^I=\sigma_1+\sigma_0 \varepsilon_S$.
Recall that $\chi=-\sigmanew_0\sigmanew_1$ is the chirality of the nut at the origin, and we could, without loss of generality,
choose the basis for $T^2$ so that $\sigmanew_0=1$ (or $\sigmanew_1=1$). 

\subsection{Example 2: \texorpdfstring{$M = \mathcal{O}(-p) \rightarrow S^2$}{M=O(-p) S2}}

We start from the toric data $\vec{v}_0=(-1,0)$, $\vec{v}_1=(0,-1)$, $\vec{v}_2=
(1,-p)$, with $d=1$, which describes the total space of the complex line bundle 
$\mathcal{O}(-p)\rightarrow S^2$. For $p=0$ this has the black hole topology $\R^2\times S^2$, while non-zero $p$ are referred to as ``Taub-bolt'' solutions. 
Using the toric data we compute
\begin{align}\label{freeR2S2}
  F_{\mathrm{grav}} = & -\frac{2}{\varepsilon_S}\bigg[\chi_S\ii\mathcal{F}(y_S^I) 
  -(1+\varepsilon_S \mskip1mu p)\chi_N\ii\mathcal{F}(y_N^I)\bigg]\Fvac\, ,
\end{align}
where, using \eqref{bweights}, $\varepsilon_S=\xi_1/\xi_2$ and $\varepsilon_N=-\varepsilon_S/(1+\varepsilon_S p)$.
Moreover, $y_N^I$, $y_S^I$ are related \eqref{yIN} by
\begin{align}
y_N^I=\frac{1}{1+\varepsilon_S\mskip1mu p}\left(y_S^I - \varepsilon_S\mskip1mu \mathfrak{p}^I\right)\, ,
\end{align}
where $\mathfrak{p}^I\equiv\mathfrak{p}^I_1$ is the magnetic flux through the 2-sphere. 
We have the constraints $\zeta_Iy_S^I=\sigma_1+\sigma_0 \varepsilon_S$ and $\zeta_I \mathfrak{p}^I=\sigma_0+\sigma_2-p\sigma_1$.
Finally, $\chi_S=-\sigma_0\sigma_1$ and $\chi_N=-\sigma_1\sigma_2$.

The free energy \eqref{freeR2S2} has a well-defined limit as $\varepsilon_S\rightarrow 0$, associated with $\xi$ having a 
bolt fixed point set,
 only when $\chi_N=\chi_S$, and in this case
\begin{align}\label{boltp}
 F_{\mathrm{grav}} & = -2\chi_S \bigg[p\mskip1mu\ii\mathcal{F}(y_S^I) 
 + \sum_{J=0}^n\mathfrak{p}^J\partial_{y_S^J}\ii\mathcal{F}(y_S^I)\bigg] \Fvac\, ,
\end{align}
which can also be obtained directly from \eqref{Fbolt}.
This correctly reproduces \footnote{We get agreement with (25) in \cite{BenettiGenolini:2024xeo} after switching to the 
$u$ variables via (20) in \cite{BenettiGenolini:2024xeo}.} the formula in \cite{BenettiGenolini:2024xeo},
which in turn gives the large $N$ field theory result of \cite{Toldo:2017qsh}. 
Thus, \eqref{freeR2S2} now generalizes this to include a rotational parameter
 $\varepsilon_S=\xi_1/\xi_2$ for the 2-sphere bolt. This should reproduce the 
large $N$ free energy of the ABJM theory on $L(p,1)=S^3/\Z_p$, now with rotational (or ``refinement'') parameter $\varepsilon_S$ for the choice of R-symmetry vector $\xi$. 

\subsection{Example 3: \texorpdfstring{$M =$ minimal resolution of $L(3,2)$}{M=minimal resolution of L(3,2)}}

We now consider the minimal resolution of the cone over $L(3,2)$.
It has two 2-sphere cycles, each with normal bundle $\mathcal{O}(-2)$ and associated magnetic fluxes $\mathfrak{p}_1^I$ and $\mathfrak{p}_2^I$. 
The toric data is given by $\vec{v}_0=(-1,0)$, $\vec{v}_1=(0,-1)$, $\vec{v}_2= (1,-2)$, $\vec{v}_3= (2,-3)$ with $d=2$.
Then we find
\begin{align}
	&\varepsilon_S
	=\frac{\xi_1}{\xi_2}\,,\qquad 
	\varepsilon_N
	=-\frac{2\xi_1+\xi_2}{3\xi_1+2\xi_2}\,. 
\end{align}
Writing
\begin{align}
\varepsilon_1\equiv\frac{b_1^1}{b_1^2}=-\frac{2\xi_1+\xi_2}{\xi_1}\,,
\end{align}
we obtain that the free energy \eqref{Fgravcute} is given by
\begin{align}
	F_{\mathrm{grav}} = & -2\bigg[\frac{\chi_S}{\varepsilon_S}\ii\mathcal{F}(y_S^I)
	+
	\frac{\chi_1}{\varepsilon_1}\ii\mathcal{F}\left(y_S^I/\varepsilon_S-\mathfrak{p}_1^I\right) 
	\nonumber\\ & 
	\qquad +\frac{\chi_N}{\varepsilon_N}\ii\mathcal{F}(y_N^I)\bigg]
	\Fvac\, .
\end{align}
From \eqref{yIN} we have
\begin{align}
y^I_N=\frac{\xi_2}{3\xi_1+2\xi_2}y^I_S-\frac{\xi_1}{3\xi_1+2\xi_2}\mathfrak{p}_1^I
-\frac{2\xi_1+\xi_2}{3\xi_1+2\xi_2}\mathfrak{p}_2^I\,,
\end{align}
 and we have the constraints $\zeta_Iy_S^I=\sigma_1+\sigma_0 \varepsilon_S$ and $\zeta_I \mathfrak{p}^I_1=\sigma_0+\sigma_2-2\sigma_1$,
 $\zeta_I \mathfrak{p}^I_2=\sigma_1+\sigma_3-2\sigma_2$.

For the special case of minimal gauged supergravity, using the procedure stated below \eqref{FSTU},
we should set $y^0_S = (\sigma_1 + \sigma_0 \varepsilon_S)/4$, $\mathfrak{p}^0_1 = (\sigma_0 + \sigma_2- 2 \sigma_1)/4$ and
$\mathfrak{p}^0_2 = (\sigma_1 + \sigma_3 - 2 \sigma_2)/4$.
If we further restrict to solutions with the chiralities all the same, $\chi_S=\chi_1=\chi_N=\pm 1$, we obtain 
(and correct \footnote{While the formula in the first line of (4.62) in \cite{BenettiGenolini:2019jdz} holds, the second line does not follow. 
Note also that we are using a different basis for the Killing vector.}) the result discussed in \cite{BenettiGenolini:2019jdz} 
 \begin{align}
 	F_{\mathrm{grav}} =\frac{3}{4} \left(2+\chi_S \frac{ -\xi_1^2+4 \xi_1 \xi_2+3 \xi_2^2}{\xi_2 (3 \xi_1+2 \xi_2)}\right)\Fvac\,.
 \end{align}
Note, in particular, that we obtain different results for $\chi_S=\pm1$.
 
\section{Extension to orbifolds}\label{sec:orbifolds}

The results we have presented have immediate extensions to 
toric 4-orbifolds, where the latter are also discussed in \cite{Calderbank:2002gy, Martelli:2023oqk, BenettiGenolini:2023yfe}. 
We assume that the supergravity fields, including the Killing spinor, 
are smooth in the appropriate orbifold sense, and that there are no additional degrees of freedom that 
contribute at the orbifold loci.

More generally now the pair of integers in $\vec{v}_a\in\Z^2$ 
do not need to be coprime: 
 the associated 2-dimensional subspace 
$D_a$, fixed by the vector $\partial_{\varphi_a}$ in \eqref{varphia}, 
has normal space $\R^2/\Z_{\mathrm{gcd}(v_a^1,v_a^2)}$.
Similarly
 $d_a=\det (\vec{v}_a,\vec{v}_{a+1})\in \Z_{>0}$, 
with $x_a$ now a local orbifold point modelled on $\R^4/\Gamma_a$, where 
$\Gamma_a$ is a finite group of order $d_a$.  
The weights~\eqref{bweights} now read
$b_{a}^1 = -\det (\vec{v}_{a+1},\vec{\xi}\mskip2mu)/d_a$, 
$b_{a}^2 = \det (\vec{v}_a,\vec{\xi}\mskip2mu)/d_a$, and 
factors of  
$1/d_a$ similarly dress various fixed point formulae, 
as explained in \cite{BenettiGenolini:2024xeo,BenettiGenolini:2024lbj} (see also \cite{BenettiGenolini:2023yfe}). Of particular note 
is that the identity \eqref{xichangebasis} still holds in the general orbifold case, 
while smoothness of the Killing spinor (in the orbifold sense) 
still implies \eqref{sigmaa} and hence \eqref{Lambdaa} also holds. 
The intersection numbers  $D_{ab}\in\mathbb{Q}$ in \eqref{Dab} are replaced with
\begin{align}\label{Daborbi}
D_{ab} = \begin{cases}\, 1/d_{a-1} & b=a- 1\\
\, -\det(\vec{v}_{a-1},\vec{v}_{a+1})/ d_{a-1} d_a & b= a\\ 
\, 1/d_{a} & b=a+ 1\\
\, 0& \mbox{otherwise}\end{cases}\, ,
\end{align}
for $1\leq a,b \leq d$ and $D_{10}\equiv 1/d_0$, $D_{d\, d+1}\equiv 1/d_d$.

For example, from the above comments one now computes the 
R-symmetry magnetic flux \eqref{Rflux} to be 
\begin{align}\label{Rfluxspindle}
\zeta_I\mathfrak{p}_a^I & = 
\frac{\sigmanew_{a-1}}{d_{a-1}}+\frac{\sigmanew_{a+1}}{d_a} +D_{aa}\sigmanew_a\, ,
\end{align}
where the first two terms on the right hand side may be recognized as the twist and anti-twist 
condition for a spindle $D_a\cong \mathbb{WCP}^1_{[d_{a-1},d_a]}$  \cite{Ferrero:2021etw} (and the last term is zero when $ND_a$ is a product). 

The free energy formula \eqref{Fgravsimp} is replaced by dividing each term in the sum with a $1/d_a$ factor. The formula
\eqref{Fgravcute} should be similarly modified as should
\eqref{UVIR}, \eqref{yRconstraint} (as illustrated in the next example).

\subsection{Example 4: \texorpdfstring{$M = \mathcal{O}(-p) \rightarrow \mathbb{WCP}^1_{[n_1,n_2]}$}{M=O(-p) over spindle}}

We consider the case that $M$ is the total space of the complex line orbibundle 
$\mathcal{O}(-p)\rightarrow\mathbb{WCP}^1_{[n_1,n_2]}$. 
Here $n_1,n_2$ are coprime positive integers, with the weighted 
projective space $\mathbb{WCP}^1_{[n_1,n_2]}$ also 
known as a spindle, and  $p\in\mathbb{Z}$. 
The following results then give a prediction for the large $N$ limit of the spindle index introduced in 
\cite{Inglese:2023wky, Inglese:2023tyc} and provide a supergravity derivation of the holomorphic block 
form of the free energy, which in the case of $p=0$ was derived in the boundary SCFT in \cite{Colombo:2024mts}. 

We define $\fg_i\equiv \mathrm{gcd}(p,n_i)$, where
the toric data specifying $M$ is $\vec{v}_0=(-\fg_1,0)$, $\vec{v}_1 = (-k, -n_1/\fg_1)$, 
$\vec{v}_2 = (q,-p/\fg_1)$.
Here $k,q\in\mathbb{Z}$ are 
any solutions to $ k p + q n_1= n_2 \fg_1$, which exist by B\'ezout's Lemma. 
The boundary $\partial M = M_3$ is, in general, an orbifold 
known as a branched lens space, but when $\mathrm{gcd}(p,q)=1=\fg_1$ it is 
the usual lens space $L(p,q)$. The above defines a broad class 
of 4-orbifolds that includes a number of special cases of interest: 
\begin{enumerate}
\item Setting $n_1=n_2=1$, then $\fg_1=1$ and we may choose $k=0$, $q=1$, which
reduces to the example $\mathcal{O}(-p)\rightarrow S^2$ studied in the previous section, 
with smooth boundary $\partial M = L(p,1)$.
\item Setting $p=1$,  again $\fg_1=1$ and we may choose $k=n_2$, $q=0$. 
This has smooth boundary $\partial M = S^3$ and corresponds 
to ``blowing up a spindle'' starting from $\R^4$ (the hyperbolic space 
${H}^4$ vacuum).
\item Setting $p=0$, so $\fg_1=n_1$, we take $k=0$, $q=n_2$, which gives 
the product space $\R^2\times \mathbb{WCP}^1_{[n_1,n_2]}$. 
This has the topology of an accelerating black hole/black saddle solution, 
with boundary $\partial M = S^1\times \mathbb{WCP}^1_{[n_1,n_2]}$.
\end{enumerate} 

We have $d_0=n_1$, $d_1=n_2$, and applying the
localization formula \eqref{Fgravsimp} (modified as stated above) we compute
\begin{align}\label{Forb}
\Fgrav = -2\Big[\frac{\chi_S}{\varepsilon_S}\ii\mathcal{F}(y_S^I) +\frac{\chi_N}{\varepsilon_N}\ii\mathcal{F}(y_N^I) \Big]\Fvac\, .
\end{align}
The $y^I_S$, $y^I_N$ variables are
as in \eqref{ys}, with the constraints
\eqref{yRconstraint} now replaced with
\begin{align}\label{yRconstraintspindle}
\zeta_Iy^I_S = \sigmanew_1 + \frac{\sigmanew_0}{n_1}\varepsilon_S\,,
\quad
 \ \zeta_Iy^I_N = \sigmanew_1 + \frac{\sigmanew_{2}}{n_2}\varepsilon_N\,, 
\end{align}
and we have defined
\begin{align}
\varepsilon_S \equiv n_1 \frac{b_0^1}{b_0^2}\, , \quad 
\varepsilon_N \equiv n_2 \frac{b_1^2}{b_1^1}\, .
\end{align}
The $y_N^I$, $y_S^I$ are related by 
\begin{align}
\frac{y^I_N}{\varepsilon_N} +\frac{y^I_S}{\varepsilon_S} = \mathfrak{p}^I\, ,
\end{align}
where $\mathfrak{p}^I\equiv\mathfrak{p}^I_1$ is the flux 
through the spindle zero-section. For example, 
in the particular case of $\R^2\times \mathbb{WCP}^1_{[n_1,n_2]}$
one can check that $\varepsilon_S=\xi_1/\xi_2=-\varepsilon_N$, 
and the above formulae reproduce those given in \cite{BenettiGenolini:2024xeo,BenettiGenolini:2024lbj}. 

The R-symmetry Killing vector just rotates the fibre of 
$\mathcal{O}(-p)$ in the limit $\varepsilon_S\rightarrow 0$ (or, equivalently, 
$\varepsilon_N\rightarrow 0$), which 
is the case
 $\vec{\xi}=(k,n_1/\fg_1)$. The 
spindle zero-section is then a bolt fixed point set in this limit. 
The formula  \eqref{Forb} has a well-defined limit only if 
$\chi_N=\chi_S$, and then
\begin{align}\label{boltorb}
 F_{\mathrm{grav}}&  = -2\chi_S \bigg[\frac{p}{n_1n_2}\ii\mathcal{F}(y_S^I) 
 + \sum_{J=0}^3\mathfrak{p}^J\partial_{y_S^J}\ii\mathcal{F}(y_S^I) \bigg] \Fvac\, .
\end{align}
This generalizes \eqref{boltp}, where $p\rightarrow p/n_1n_2$ simply replaces 
the self-intersection number 
of the bolt by its orbifold generalization. The fluxes 
$\mathfrak{p}^I$ satisfy the constraint
\begin{align}\label{orbiflux}
\zeta_I \mathfrak{p}^I = \frac{\sigmanew_0}{n_1} + \frac{\sigmanew_2}{n_2} -\frac{p}{n_1n_2}\sigmanew_1\, ,
\end{align}
where the bolt limit requires us to set $\sigmanew_0=\sigmanew_2$,
while the constraint reads $\zeta_Iy^I_S = \sigmanew_1$, (and recall the chirality is $\chi=-\sigmanew_0\sigmanew_1$). 

The special case of \eqref{boltorb} where $p=1$ 
gives a smooth boundary $\partial M=S^3$, and the 
R-symmetry Killing vector that gives a spindle bolt 
in the bulk is simply $\xi=(n_2,n_1)$. This 
is also clear from the fact that 
$\mathbb{WCP}^1_{[n_1,n_2]}=S^3/U(1)_R$, 
where $\xi$ generates $U(1)_R$. In particular, 
we expect that the large $N$ limit of the ABJM theory 
on $S^3$, with R-symmetry Killing vector 
$\xi=(n_2,n_1)$, should exhibit saddle points 
with free energy \eqref{boltorb} with $p=1$ and STU prepotential \eqref{FSTU}, 
generalizing 
the analysis of the 2-sphere case of \cite{Toldo:2017qsh} 
to a spindle. It is already known that saddle points 
exist for the filling $M=\R^4$ with the same R-symmetry 
vector~\cite{Bobev:2022eus}.

\section{Discussion}

We have presented general formulae for computing the gravitational free energy of toric gravitational instantons
of $\mathcal{N}=2$ gauged supergravity coupled to vector multiplets. When the $D=4$ theory arises as a consistent 
KK truncation of $D=10,11$ supergravity, we immediately obtain results for the higher-dimensional gravity solutions 
and hence the corresponding dual SCFTs. For example,
solutions of the STU model can be uplifted on $S^7$ \cite {Cvetic:1999xp} to obtain solutions dual to ABJM theory, 
while solutions of minimal gauged supergravity can also be uplifted on $SE_7$ and other manifolds to $D=11$ \cite {Gauntlett:2007ma}
giving solutions dual to other SCFTs. 
However, we believe our results have broader applicability,
and that even when such a consistent KK truncation does not exist, the corresponding gauged supergravity sector we study is 
still sufficient to evaluate the gravitational free energy, at least in some cases. Indeed our result
\eqref{boltp} is in precise agreement with the field theory result in \cite{Toldo:2017qsh}, in the large $N$ limit, for the classes of $\mathcal{N}=2$
SCFTs considered there.
In making this comparison, one should identify the prepotential of the gauged supergravity with the twisted superpotential
or Bethe potential arising in the field theory computation, a point that was 
originally made in \cite{Hosseini:2016tor}.

From a holographic perspective, we have considered solutions that are associated with the dual SCFT 
placed on a branched lens space, the most general toric 3-dimensional orbifold space. 
In the bulk it is natural to partially resolve the orbifold singularities by blowing up a spindle.
However, it is also possible to further resolve by considering additional blow ups, adding more 2-cycles, eventually
leading to a completely regular solution with a number of $S^2$ 2-cycles. 
In all cases, we have shown how to simply compute the gravitational free energy, provided that the corresponding solution actually exists.

For the specific case of a single 2-cycle in the bulk we have examined some particular cases. We
have generalized the case of a lens space $L(p,1)$ boundary with an $S^2$ bolt, for which specific solutions were found in \cite{Toldo:2017qsh}, to 
also include a rotation/refinement parameter $\varepsilon$ in the choice of R-symmetry vector.
In addition, for $S^3$ boundary we have also considered the possibility of solutions with a spindle $\mathbb{WCP}^1_{[n_1,n_2]}$ bolt 
where the R-symmetry vector is simply $\xi=(n_2,n_1)$. Equations \eqref{freeR2S2}, and \eqref{boltorb} with $p=1$ 
give gravity predictions for the large $N$ free energy of such 
saddle points, respectively.

Our results allow one to obtain the gravitational free energy for the general class of toric gravitational instantons that we have studied, provided that the solutions exist. Proving existence theorems is left as an interesting and important open problem in PDE's. While we anticipate some restrictions, we also anticipate that a very rich set of solutions exists 
and hence there is a corresponding rich class of saddle points to be discovered in the dual SCFTs in the large $N$ limit.

\section*{Acknowledgments}
This work is supported  by STFC grants ST/X000575/1 and
ST/T000864/1, EPSRC grant EP/R014604/1, and SNSF Ambizione grant PZ00P2\_208666.
PBG thanks the UCSB Physics Department for hospitality. 
JPG is a Visiting Fellow at Perimeter Institute. 
AL is supported by a Palmer Scholarship.

\bibliography{biblio}{}


\vspace{2cm}

\appendix

\section{Supplementary material: Bolts from nuts}
The localization formula for the gravitational free energy \eqref{Fgravsimp}
is valid when all fixed points of $\xi$ 
are isolated, meaning that $b_a^i\neq 0$ for all $a=0,\ldots,d$. 
If we fix $a=a'\in \{1,\ldots,d\}$ and consider the ``bolt limit" $b_{a'}^2=-b_{a'-1}^1\rightarrow 0$,
the 2-sphere $D_{a'}$ becomes a fixed point set of $\xi$ (i.e. a bolt) 
with the other fixed points remaining isolated. 
Since the Killing spinor $\epsilon$ necessarily
has fixed chirality over a connected component of the fixed point 
set of $\xi$ \cite{BenettiGenolini:2024xeo,BenettiGenolini:2024lbj}, we must 
have $\chi_{a'}=\chi_{a'-1}$, which implies $\sigmanew_{a'-1}=\sigmanew_{a'+1}$.
Now from \eqref{pflux} we have
\begin{align}
\Lambda_{a'-1}^I=\Lambda^I_{a'}-2b^2_{a'}\mathfrak{p}^I_{a'}\,,
\end{align}
which shows that in the bolt limit 
we also have $\Lambda^I_{a'-1}=\Lambda^I_{a'}$; in fact this had to be the case 
since $\Lambda^I$ is necessarily constant over a connected component of the fixed 
point set of $\xi$ (here $D_{a'}$) \cite{BenettiGenolini:2024xeo,BenettiGenolini:2024lbj}. Using the constraints on the toric vectors and using \eqref{Dab},
one can also show that
\begin{align}
b^2_{a'-1}=b^1_{a'}-b^2_{a'}D_{a'a'}\,,
\end{align}
and hence in the bolt limit we have $b^2_{a'-1}=b^1_{a'}$.
We then find that \eqref{Fgravsimp} is well-defined in the bolt limit, leading to
\begin{align}
\label{Fbolt}
	&F_{\mathrm{grav}} =  -\frac{\pi}{G_N}\Bigg[\sum_{\substack{a=0 \\ a\neq a',\mskip2mu a'-1}}^d \frac{\chi_a}{4b_a^1b_a^2}\ii \mathcal{F}(\Lambda^I_a) \nonumber\\
	& -\frac{\chi_{a'}}{4(b^1_{a'})^2}D_{a'a'}\ii \mathcal{F}(\Lambda^I_{a'})+
	\frac{\chi_{a'}}{2b^1_{a'}}\sum_{J=0}^n\ii \mathcal{F}_J(\Lambda^I_{a'})\mathfrak{p}_{a'}^J\Bigg]\, ,
\end{align}
where $\mathcal{F}_J(X^I)\equiv \partial_{X_J}\mathcal{F}(X^I)$ denotes 
a partial derivative with respect to the argument. 

This can also be written 
\begin{align}\label{Fbolt2}
	&F_{\mathrm{grav}} =  -\frac{\pi}{G_N}\Bigg[\sum_{\substack{a=0 \\ a\neq a',\mskip2mu a'-1}}^d 
	\chi_a \frac{(b_a^1- \chi_a b_a^2)^2}{b_a^1b_a^2}\ii \mathcal{F}\Big(\frac{\Lambda^I_a}{\Lambda_a}\Big)
	 \nonumber\\
	& -{\chi_{a'}}D_{a'a'}\ii \mathcal{F}\Big(\frac{\Lambda^I_{a'}}{\Lambda_{a'}}\Big)-\sigma_{a'-1}\sum_{J=0}^n\ii \mathcal{F}_J\Big(\frac{\Lambda^I_{a'}}{\Lambda_{a'}}\Big)\mathfrak{p}_{a'}^J\Bigg]\, ,
\end{align}
using the fact that $\mathcal{F}$ is homogeneous of degree two.
The second line of
\eqref{Fbolt2} agrees with the general formula for a bolt 
contribution in \cite{BenettiGenolini:2024xeo,BenettiGenolini:2024lbj}, showing 
self-consistency of the general fixed point formula. The first line, present for $d>1$, gives the contribution
of the remaining isolated fixed points. One can treat multiple bolts similarly.

Finally, we note that the limits $b_0^2\rightarrow 0$, $b_1^d\rightarrow 0$, not considered above,
correspond to $D_0$ and $D_{d+1}$ being fixed under 
$\xi$, respectively, which we have tacitly assumed is not the case in writing 
\eqref{ys}. In this limit $\xi$ would have a fixed point locus on $\partial M$.

\end{document}